\documentclass[aps,pra,twocolumn,reprint,floatfix,longbibliography]{revtex4}
\usepackage[utf8]{inputenc} 
\usepackage{graphicx}
\usepackage{amsmath}
\usepackage{amsfonts}
\usepackage{mathtools}
\usepackage{float}
\usepackage{tabularx, booktabs}
\usepackage{color}
\usepackage{braket}
\usepackage{bm}
\usepackage{comment}

\usepackage{hyperref}

\hypersetup{
	colorlinks=true,
	linkcolor=blue,
	citecolor=blue,
	filecolor=magenta,
	urlcolor=cyan,
	breaklinks=true
}
\newcommand{\papertitle}{Binding of heavy fermions by a single light atom in one dimension}
\newcommand{\lptms}{Universit\'e Paris-Saclay, CNRS, LPTMS, 91405 Orsay, France}

\DeclareSymbolFont{sfletters}{OML}{cmbrm}{m}{it}

\begin{document}
	
\title{\papertitle}
\author{A. Tononi}
\affiliation{\lptms}
\author{J. Givois}
\affiliation{\lptms}
\author{D. S. Petrov}
\affiliation{\lptms}

\date{\today}

\begin{abstract}
We consider the problem of $N$ identical fermions interacting via a zero-range attractive potential with a lighter atom in one dimension. Using the few-body approach based on the Skorniakov and Ter-Martirosian equation, we determine the energies and the critical mass ratios for the emergence of the tetramer, pentamer, and hexamer. For large $N$, we solve the problem analytically by using the mean-field theory based on the Thomas-Fermi approximation. The system becomes bound when the heavy-to-light mass ratio exceeds a critical value which grows as $N^3$ at large $N$. We also employ a more sophisticated Hartree-Fock approach, which turns out to be equivalent to the Thomas-Fermi approximation for determining the energies, but provides a better description of the microscopic structure of the clusters.

\end{abstract}

\maketitle

The binding of heavy fermions by a light atom is a phenomenon which originates from the competition between the Pauli pressure for the heavy fermions and the induced attraction due to the exchange of the light atom. 
This problem, in the case of zero-range interactions in free space, is parametrized by the mass ratio $M/m$, the number of heavy fermions $N$, and spatial dimensionality. The heavy-light scattering length $a>0$, which determines the size of the $1+1$ molecule, is the only dimensional parameter, and can be used as the unit of length. The $2+1$ trimer case is well understood in any dimension: in three-dimensions the trimer emerges for $M/m>8.173$~\cite{KartavtsevMalykh}, in two dimensions for $M/m>3.33$~\cite{PricoupenkoPedri}, and in one dimension for $M/m>1$~\cite{Kartavtsev}. The reduction of the critical mass ratio is explained by the qualitative argument that the centrifugal barrier for the heavy fermions effectively lowers when decreasing the dimension. 

The $N+1$ cluster problem was discussed for $N\leq 4$ both in three~\cite{Blume,bazak2017} and in two dimensions~\cite{levinsen,liu2022}. It turns out that, with increasing the mass ratio, it becomes possible to bind more and more heavy atoms. Treating higher $N$ is very challenging as the problem is nonperturbative, featuring the fermionic sign problem, significant shell effects, and the onset of the Efimov effect in three dimensions at $M/m\approx 13$ \cite{Efimov,Castin,bazak2017}. In one dimension these problems either do not exist or are less pronounced. Thus, this geometry seems suitable for connecting the limits of small and large $N$, and for better understanding the binding mechanism. The largest free-space cluster in one dimension, the $3+1$ tetramer, was studied by Mehta~\cite{mehta2014} in the Born-Oppenheimer approximation, valid for $M\gg m$.

\begin{figure}[hbtp]
\includegraphics[width=0.99\columnwidth]{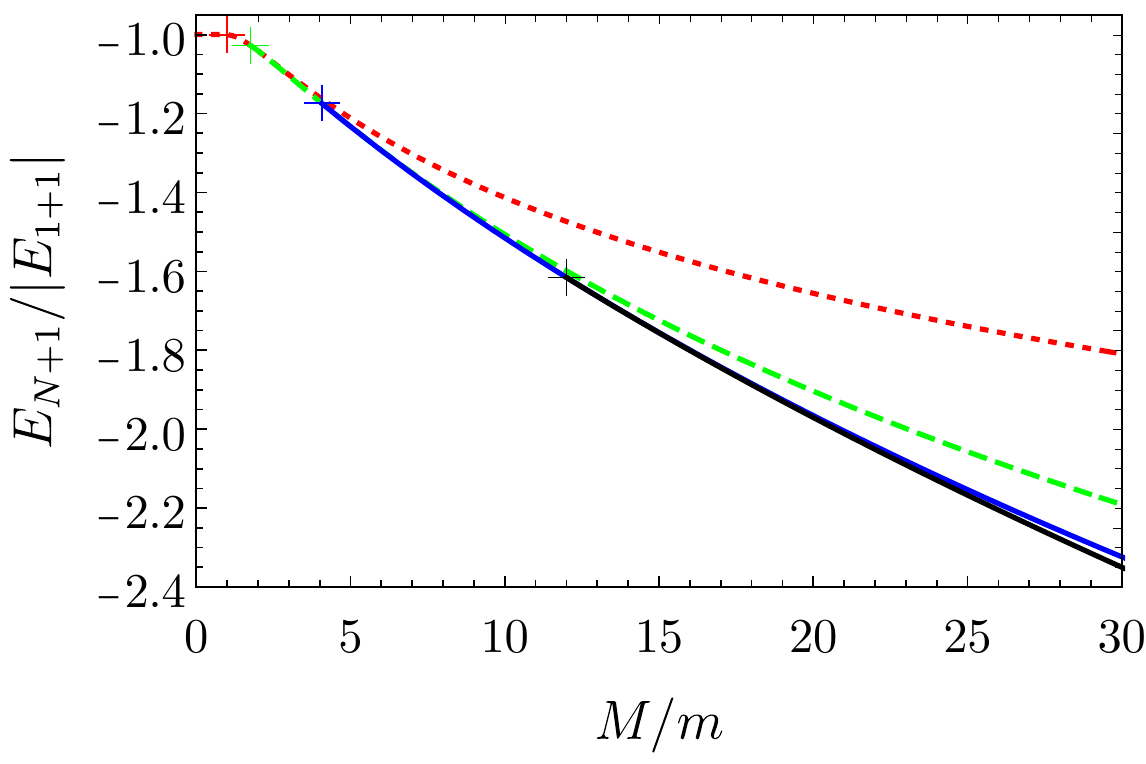}
\caption{
Bound-state energies of the trimer (red dotted), tetramer (green dashed), pentamer (blue solid), and hexamer (black solid) versus the mass ratio ${M/m}$. The crosses indicate the thresholds for the corresponding clusters.
}
\label{fig1}
\end{figure}

Here we solve the one-dimensional (1D) $N+1$ cluster problem exactly for $N$ up to 5, determining the corresponding critical mass ratios (see Fig.~\ref{fig1}). Then, using the local Thomas-Fermi approximation for the kinetic energy of the heavy fermions and the mean-field assumption for the interaction, both valid in the large $N$ limit, we predict that the $N+1$ cluster binds at $M/m=\pi^2N^3/36$, and we describe its properties (size, shape, energy) analytically. Aiming to improve the accuracy and the description of the clusters with moderate $N$, we turn to the Hartree-Fock method. We find that both the Thomas-Fermi and Hartree-Fock approaches predict essentially the same energies for the clusters. However, the latter gives more details on the structure, and well reproduces momentum correlations. The energy determination with this method can be significantly improved by taking higher Hartree-Fock orbitals into account in a perturbative manner. The mean-field methods that we employ in this paper are similar to the ones used for studying bright solitons and liquid droplets in 1D Bose-Fermi mixtures~\cite{Karpiuk2004,Rakshit2019}.

Our mixture is described by the Hamiltonian 
\begin{equation}\label{Ham}
H=\int\left( -\frac{\hat{\Psi}^\dagger_x \partial^2_x\hat{\Psi}_x}{2M}-\frac{\hat{\phi}^\dagger_x \partial^2_x\hat{\phi}_x}{2m}+g\hat{\Psi}^\dagger_x \hat{\phi}^\dagger_x \hat{\Psi}_x\hat{\phi}_x\right)dx,
\end{equation}
where $\hat{\Psi}^\dagger_x$ and $\hat{\phi}^\dagger_x$ are the creation operators of, respectively, heavy and light fermions at position $x$. We set $\hbar=1$, and assume an attractive light-heavy interaction characterized by $g=-1/(m_r a)<0$ and by the dimer binding energy $E_{1+1} = - 1/(2 m_r a^2)$, where $m_r=Mm/(M+m)$.

The $N+1$-body problem, following the method of Skorniakov and Ter-Martirosian (STM)~\cite{stm}, who considered the $2+1$ case in three dimensions, can be reformulated as an integral equation for a function $F(q_1,...,q_{N-1},q_N)$, which is the Fourier transform of the total $N+1$-body wave function $\Psi(x_1,...,x_{N-1},x_N,x=x_N)$, where the light atom coordinate $x$ is set to coincide with $x_N$. 
The function $F$ describes $N-1$ heavy atoms with momenta $q_1,...q_{N-1}$ plus a heavy-light pair with momentum $q_N$. By setting the (conserved) total momentum to zero, the pair momentum $q_N$ is replaced by $-\sum_{i=1}^{N-1} q_i$, and can be omitted. The resulting STM equation reads~\cite{pricoupenko2011,pricoupenko2018} 
\begin{align}
&\bigg[ \frac{a}{2} - \frac{1}{2\kappa(q_1,...,q_{N-1})}
\bigg] F(q_1,...,q_{N-1}) = 
\nonumber
\\
&- \int \frac{dp}{2\pi} \frac{\sum_{j=1}^{N-1} F(q_1,...,q_{j-1},p,q_{j+1},...,q_{N-1})}{\kappa^2(q_1,...,q_{N-1})+(p+\frac{m_r}{m} \sum_{i=1}^{N-1} q_i)^2}, 
\label{stmeq}
\end{align}
where $\kappa(q_1,...q_{N-1})=\sqrt{-2m_r E+\frac{m_r}{M+m} (\sum_{i=1}^{N-1} q_i)^2+\frac{m_r}{M} \sum_{i=1}^{N-1} q_i^2}$ and $E$ is the total energy, assumed negative. 

To facilitate the numerical solution 
of Eq.~(\ref{stmeq}) we restrict the space of the problem to ordered configurations $q_1<q_2<...<q_{N-1}$ and use the antisymmetry property $F(q_1,...,q_{N-1}) = s F(\bar{q}_1,...,\bar{q}_{N-1})$, where $(\bar{q}_1,...,\bar{q}_{N-1})$ is the correctly ordered permutation of $(q_1,...,q_{N-1})$ and $s$ is the sign of the ordering permutation. Moreover, since Eq.~(\ref{stmeq}) conserves parity, i.e., $F(q_1,...,q_{N-1})=P F(-q_1,...,-q_{N-1})$ with $P=\pm1$, we treat the $P=+1$ and $P=-1$ sectors separately, restricting the configurational space to configurations with $\sum_{i=1}^{N-1} q_i\ge 0$. We obtain the solution of Eq.~(\ref{stmeq}) by discretizing the momenta and diagonalizing the resulting matrix, finding, for any mass ratio $M/m$, the value of $a$ versus $E$.

Our results for the bound-state energies up to the $5+1$ hexamer are shown in Fig.~\ref{fig1}. The critical mass ratios for the emergence of the trimer, tetramer, pentamer, and hexamer are marked by crosses. They equal, respectively, $(M/m)_{2+1}=1$ (see Ref.~\cite{Kartavtsev}), $(M/m)_{3+1}=1.76$, $(M/m)_{4+1}=4.2$ and $(M/m)_{5+1}=12.0$, the last of these being determined with about 5\% accuracy. The trimer and tetramer energies are in agreement, respectively, with Ref.~\cite{Kartavtsev} and, asymptotically, with the Born-Oppenheimer result of Mehta~\cite{mehta2014}. The parity of the trimer and tetramer is negative, in accordance with Refs.~\cite{Kartavtsev,mehta2014}, while the pentamer and hexamer are characterized by $P=+1$. 

We now approach the $N+1$-cluster problem from the limit of large $N$ employing two different mean-field methods (cf. Refs.~\cite{Karpiuk2004,Rakshit2019}). The first is based on the Thomas-Fermi approximation for the heavy fermions. Namely, we minimize the grand potential
\begin{eqnarray}\label{DF}
\Omega=&\int [ |\phi'(x)|^2/2m+gn(x)|\phi(x)|^2+\pi^2n^3(x)/6M\nonumber\\
&-\epsilon |\phi(x)|^2-\mu n(x)]\, dx,
\end{eqnarray}
where $\phi(x)$ is the wave function of the light atom and $n(x)$ is the density of the heavy atoms. The term proportional to $n^3$ in Eq.~(\ref{DF}) is the Thomas-Fermi kinetic energy density of the heavy fermions in one dimension. The Lagrange multipliers $\epsilon$ and $\mu$ account, respectively, for the normalization conditions
\begin{align}\label{NormLight}
\int |\phi(x)|^2 \, dx = 1,
\\
\label{NormHeavy}
\int n(x) \, dx = N.
\end{align}

The minimization of $\Omega$ with respect to $n(x)$ gives
\begin{equation}\label{nx}
n(x)=\left\{
 \begin{array}{cc}
  \sqrt{-2Mg[|\phi(x)|^2-\mu/g]}/\pi, &  |\phi(x)|^2>\mu/g \\
  0, & |\phi(x)|^2\leq \mu/g.
 \end{array}\right.
 \end{equation}
Then, to obtain the wave function of the light atom  $\phi(x)$, we restrict our analysis to $\mu< 0$ and $\epsilon<0$ (since we are interested in bound states), take $\phi$ to be real, and assume that it has a symmetric bell shape, centered at the origin. We then introduce the Thomas-Fermi size $x_{\rm TF}$ such that $\phi^2(x_{\rm TF})=\mu/g$: the function $n(x)$ is thus nonzero only in the interval $(-x_{\rm TF},x_{\rm TF})$.
The minimization of Eq.~(\ref{DF}) with respect to $\phi$ inside this interval leads to the equation
\begin{equation}\label{Schr}
-\phi''(x)-\frac{\sqrt{-8Mm^2g^3}}{\pi}\sqrt{\phi^2(x)-\mu/g}\phi(x)=2m\epsilon\phi(x),
\end{equation}  
the solution of which should be matched with the free solution $\phi(x)=\phi(x_{\rm TF})\exp[-\sqrt{-2m\epsilon}(|x|-x_{\rm TF})]$, valid for $|x|>x_{\rm TF}$. Multiplying Eq.~(\ref{Schr}) by $\phi'$, and integrating over $x$, we get
\begin{equation}\label{FirstIntegral}
\phi'^2=-2m\epsilon \phi^2 -\frac{\sqrt{-32Mm^2g^3}}{3\pi}(\phi^2-\mu/g)^{3/2}.
\end{equation}

The condition $\phi'(0)=0$ [since we assume $\phi(x)=\phi(-x)$], substituted into Eq.~(\ref{FirstIntegral}), implies
\begin{equation}\label{phiTFphi0}
\phi^2(x_{\rm TF})=\mu/g=\phi^2(0)(1-\lambda),
\end{equation}
where we introduced 
\begin{equation}\label{Lambda}
\lambda=\left[\frac{9\pi^2\epsilon^2}{-8Mg^3\phi^2(0)}\right]^{1/3}.
\end{equation} 
This dimensionless parameter ranges from 0 to 1, and characterizes the shape of the droplet [see Eq.~(\ref{phiImplicit})]. The limit $\lambda \rightarrow 1$ indicates the binding threshold for a new heavy atom, since in this case $\mu=0$. In this limit, Eq.~(\ref{FirstIntegral}) can be integrated analytically, resulting in 
\begin{equation}\label{phianalytic}
\phi(x)=\frac{-3\pi \epsilon}{\sqrt{-8Mg^3}}\frac{1}{\cosh^2(\sqrt{-m\epsilon/2}x)}.
\end{equation}
Imposing the normalization conditions Eqs.~(\ref{NormLight}) and (\ref{NormHeavy}) on $\phi(x)$ we obtain that this limit, i.e., the threshold for the binding of $N$ heavy atoms~\cite{remTF}, is realized for 
\begin{equation}\label{Mratiothreshold}
\frac{M}{m}=\frac{\pi^2}{36}N^3,
\end{equation}
the size of the cluster at this point is given by
\begin{equation}\label{Sizethreshold}
\frac{1}{\sqrt{-2m\epsilon}}=\frac{3}{-mgN},
\end{equation}
and the total energy reads
\begin{equation}\label{Energythreshold}
E=\epsilon+\int \frac{\pi^2n^3(x)}{6M}\,  dx=-\frac{mg^2}{30}N^2=-\frac{mg^2}{30}\left(\frac{36}{\pi^2}\frac{M}{m}\right)^{2/3}.
\end{equation}
We see that for large $N$, which is the validity criterion of the Thomas-Fermi approximation, the mean-field treatment of the light-heavy interaction is also justified, since the typical de Broglie wave length of the atoms is much smaller than the scattering length $a=-1/(m_rg)\approx -1/(mg)$. 

More generally, for $\lambda<1$, i.e., when $M/m>\pi^2N^3/36$, various cluster properties can be found by writing $\int_0^x{\cal F}[\phi(x)]dx=\int_\phi^{\phi(0)} {\cal F}[\phi(x)]d\phi/|\phi'|$, valid in the interval $0<x<x_{\rm TF}$. In particular, setting ${\cal F}=1$ and using $\phi'$ from Eq.~(\ref{FirstIntegral}), we obtain the dependence $x(\phi)$ [inverse of $\phi(x)$]: 
\begin{equation}\label{phiImplicit}
\sqrt{-2m\epsilon}x=\lambda^{3/4}\int_{\phi/\phi(0)}^1\frac{dy}{\sqrt{\lambda^{3/2}y^2-(y^2-1+\lambda)^{3/2}}}.
\end{equation}
Similarly, by using ${\cal F}(\phi)=\phi^2$ and ${\cal F}(\phi)=\sqrt{\phi^2-\phi^2(x_{\rm TF})}$, Eqs.~(\ref{NormLight}) and (\ref{NormHeavy}) can be rewritten, respectively, as
\begin{equation}\label{NormLightRescaled}
\sqrt{\frac{3\pi}{4\sqrt{2}}}\frac{\phi^{3/2}(0)}{(-mg)^{3/4}}\left(\frac{M}{m}\right)^{-1/4}I_l(\lambda)=1
\end{equation}
and
\begin{equation}\label{NormHeavyRescaled}
\sqrt{\frac{3}{2\sqrt{2}\pi}}\frac{\phi^{1/2}(0)}{(-mg)^{1/4}}\left(\frac{M}{m}\right)^{1/4}I_h(\lambda)=N,
\end{equation}
with
\begin{equation}\label{Il}
I_l(\lambda)=\frac{1-\lambda}{\lambda^{3/4}}+2\int_{\sqrt{1-\lambda}}^1\frac{y^2 \, dy}{\sqrt{\lambda^{3/2}y^2-(y^2-1+\lambda)^{3/2}}}
\end{equation}
and
\begin{equation}\label{Ih}
I_h(\lambda)=2\int_{\sqrt{1-\lambda}}^1\frac{\sqrt{y^2-1+\lambda} \, dy}{\sqrt{\lambda^{3/2}y^2-(y^2-1+\lambda)^{3/2}}}.
\end{equation}
The total energy of the cluster equals
\begin{equation}\label{EnergyRescaled}
E = \epsilon+\frac{1}{2^{3/4}\sqrt{3\pi}}\frac{\phi^{5/2}(0)(-mg)^{3/4}}{m}\left(\frac{M}{m}\right)^{1/4}I_E(\lambda),
\end{equation}
where
\begin{equation}\label{IE}
I_E(\lambda)=2\int_{\sqrt{1-\lambda}}^1\frac{(y^2-1+\lambda)^{3/2} \, dy}{\sqrt{\lambda^{3/2}y^2-(y^2-1+\lambda)^{3/2}}}.
\end{equation}
Equations~(\ref{NormLightRescaled}) and (\ref{NormHeavyRescaled}) can be solved for $\phi(0)$ and $\lambda$ as functions of $N$, $M$, $m$, and $g$. Then, the shape of the cluster follows from Eq.~(\ref{phiImplicit}), and its total energy from Eq.~(\ref{EnergyRescaled}), where $\epsilon$ is deduced from Eq.~(\ref{Lambda}). 

We note that $\lambda$ depends only on the combination $mN^3/M$. Indeed, by eliminating $\phi(0)$ from Eqs.~(\ref{NormLightRescaled}) and (\ref{NormHeavyRescaled}), one obtains the implicit equation for $\lambda$
\begin{equation}\label{lambdaSol}
\frac{M}{m}=\frac{2\pi^2}{3}\frac{I_l(\lambda)}{I_h^3(\lambda)}N^3.
\end{equation}
This means that two clusters with different $N$ and $M/m$ but equal $mN^3/M$ have similar shapes. This is valid as long as the de Broglie wave lengths of the atoms are much smaller than $a$, in our case equivalent to $N\gg 1$. 

Finally, let us discuss the limit $\lambda\rightarrow 0$, reached when $M/m\gg N^3$. In this case, it is convenient to change the integration variable in Eqs.~(\ref{phiImplicit}), (\ref{Il}), (\ref{Ih}), and (\ref{IE}) to $t=(y-\sqrt{1-\lambda})/(1-\sqrt{1-\lambda}$), such that the integrals can be expanded at small $\lambda$. 
At the leading order we obtain
$x_{\rm TF}\sqrt{-2m\epsilon}\approx \lambda\sqrt{\pi}\Gamma(5/3)/2\Gamma(7/6)\ll 1$, $\lambda\approx (9\pi^2N^3m/32M)^{1/3}$, $\phi(0)\approx \sqrt{-mgN}$, and $E\approx\epsilon\approx -N^2mg^2/2$. We see that the heavy atoms are much more localized than the light atom, and, therefore, we deal with a halo bound state. We observe that the total energy of the $N+1$ cluster in this limit is dominated by $\epsilon$, the kinetic energy of the heavy atoms being subdominant. We also note that this theory predicts that $E_{N+1}\rightarrow N^2 E_{1+1}$ when $M/m\rightarrow \infty$, which is the exact asymptote for any $N$. Indeed, in this limit, the heavy atoms are easily localized in a narrow spatial interval, and the problem reduces to the scattering of a light atom by a fixed potential $Ng\delta(x)$~\cite{RemLargeMratio}.

\begin{figure}[hbtp]
\includegraphics[width=0.99\columnwidth]{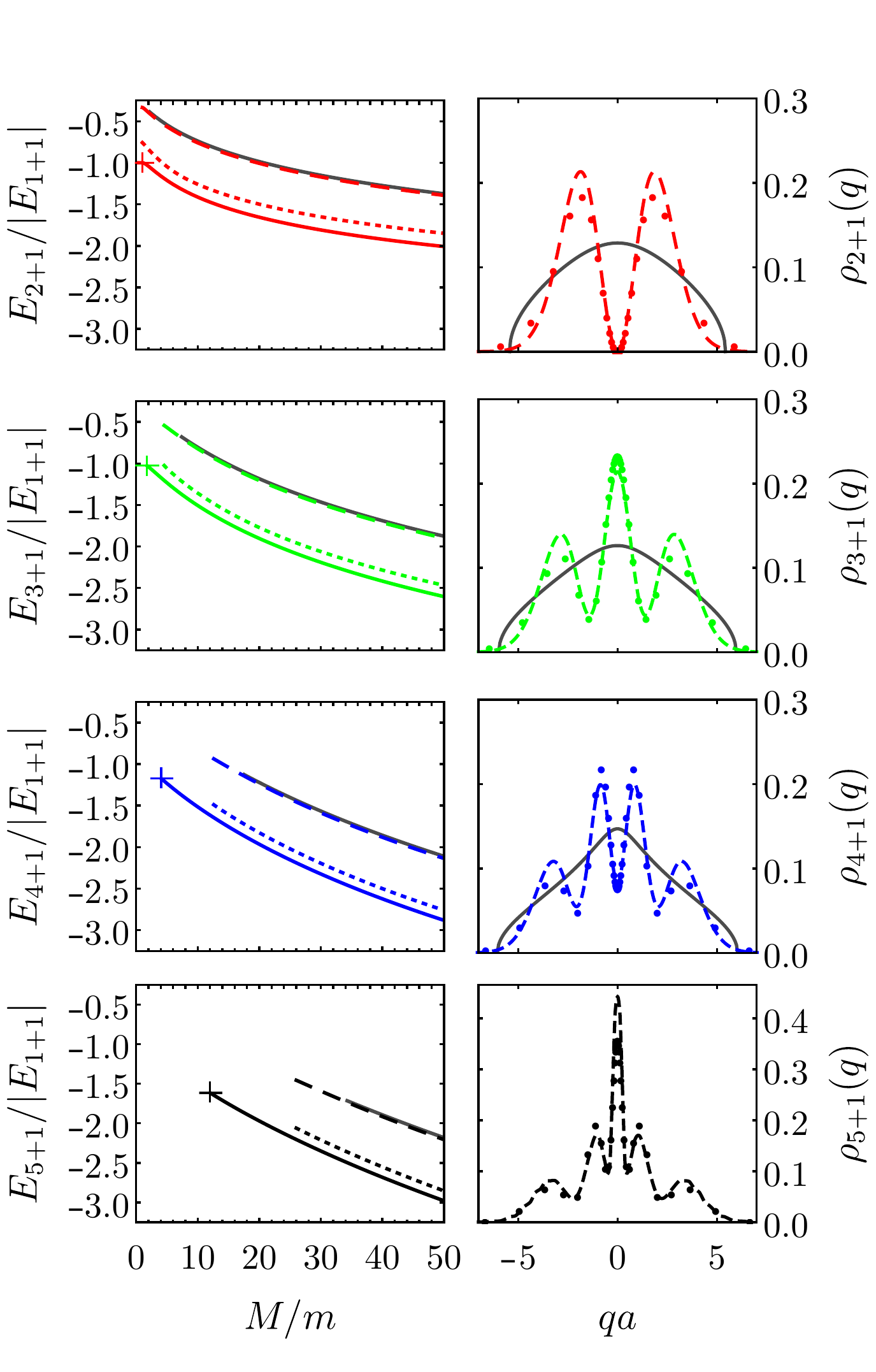}
\caption{
Left column: cluster energies for $N$ up to 5 obtained by various methods. Solid curves correspond to the exact result, solid gray to the prediction of the Thomas-Fermi method, dashed to the Hartree-Fock approximation, and dotted to the Hartree-Fock result with the second-order correction (see text). Right column: $\rho_{N+1}(q)$ normalized to 1 for the $N+1$ clusters at $M/m=28.75$, corresponding to the ${}^{173}\textrm{Yb}$--${}^{6}\textrm{Li}$ mixture. The dots correspond to the exact solution, gray curves to the Thomas-Fermi method, and dashed curves to the Hartree-Fock method. The hexamer is not bound in the Thomas-Fermi approach for the chosen mass ratio.
}
\label{fig2}
\end{figure}

In the left panels of Fig.~\ref{fig2} we compare the cluster energies calculated by using the Thomas-Fermi approximation (highest gray solid curves) with the exact results (lowest solid curves, same data as in Fig.~\ref{fig1}). We see that, although the Thomas-Fermi approximation provides a clear physical understanding and a simple (even analytical) treatment of the problem, it is not very precise at these finite $N$.  The discrepancy comes from the local-density approximation for the heavy fermions and from the breakdown of the mean-field approximation for the interaction. 

To better understand which one of these factors is dominant, and to improve the theory, we turn to another mean-field approach based on the Hartree-Fock variational method. Within this method, we look for the variational estimate of the ground-state energy of the Hamiltonian~(\ref{Ham}) by using the variational ansatz
\begin{equation}\label{VarWF}
\ket{v}=\int dx\phi_1(x)\hat{\phi}^\dagger_x\int dx_1...dx_N\frac{\det[\Psi_\nu(x_\eta)]}{\sqrt{N!}}\prod_{\eta=1}^N\hat{\Psi}^\dagger_{x_\eta}\ket{0}. 
\end{equation}
The minimization of the variational energy $\bra{v}H\ket{v}$ with respect to the orbitals $\phi_1$ and $\Psi_\nu$ ($\nu=1,...,N$), subject to the normalization constraints, gives the set of $N+1$ equations
\begin{eqnarray}
&-\partial^2_x\phi_1(x)/2m+gn(x)\phi_1(x)=\epsilon_1\phi_1(x),\label{SchrLight}\\
&-\partial^2_x\Psi_\nu(x)/2M+g|\phi_1(x)|^2\Psi_\nu(x)=E_\nu\Psi_\nu(x),\label{SchrHeavy}
\end{eqnarray}
where $n(x)=\sum_{\nu=1}^N|\Psi_\nu(x)|^2$, and all the orbital wave functions are normalized. The Hartree-Fock procedure is very simple in our case, since the heavy fermions do not interact. We solve Eqs.~(\ref{SchrLight}) and (\ref{SchrHeavy}) iteratively. Namely, we pick an initial guess for $\phi_1(x)$ [we use Eq.~(\ref{phianalytic})], diagonalize Eq.~(\ref{SchrHeavy}), plug its $N$ lowest (normalized) eigenstates into Eq.~(\ref{SchrLight}), and solve for $\phi_1$. This procedure is repeated until convergence. The variational energy is given by
\begin{equation}\label{VarEn}
\bra{v}H\ket{v}=\epsilon_1+\sum_{\nu=1}^N E_\nu-g\int |\phi_1(x)|^2n(x) \, dx.
\end{equation}

The cluster energies calculated by this procedure are shown as dashed curves in the left panels of Fig.~\ref{fig2}. We see that the Hartree-Fock and Thomas-Fermi approaches give very close results, and it seems that the first does not bring any improvements over the second, even though it does not rely on the local-density assumption for the kinetic energy. However, the Hartree-Fock method provides us with the wave function and gives us an access to the interparticle correlations, which can be compared to the exact solution.

As a measure of these correlations, easily extracted from all the considered methods, we take the quantity $\rho_{N+1}(q) = \int |F(q,q_2,...,q_{N-1})|^2 \, dq_2 ... dq_{N-1}$, which is the momentum distribution of the remaining $N-1$ heavy atoms once a heavy-light pair is instantly quenched and removed from the system (for instance, by using the photoassociation technique). In the Thomas-Fermi approximation, this quantity is 
the momentum distribution of an ideal Fermi sea of density $n(x)$, i.e., flat for $-\pi n(x)<q<\pi n(x)$, averaged over $x$. For $q>0$ it is proportional to $x(q/\pi)$, where $x(n)$ is the inverse of $n(x)$, and is given by Eq.~(\ref{phiImplicit}) with $\phi$ expressed through $n$ by Eq.~(\ref{nx}). In the right panels of Fig.~\ref{fig2} we show $\rho_{N+1}(q)$ extracted from the STM equation (dots), from the Hartree-Fock wave function (dashed curves), and from the Thomas-Fermi $n(x)$ (gray solid curves) for various clusters with $M/m=28.75$, corresponding to the ${}^{173}\textrm{Yb}$--${}^{6}\textrm{Li}$ atomic mixture. All the curves are normalized to $1$. 

We see that the Hartree-Fock approach well approximates $\rho_{N+1}(q)$, which suggests that this method provides a good starting point for a more precise energy determination in a computationally inexpensive manner. We note that the linear Schr\"odinger operators on the left-hand sides of Eqs.~(\ref{SchrLight}) and (\ref{SchrHeavy}), with previously determined $n(x)$ and $|\phi_1(x)|^2$ playing the roles of potentials, give us orthonormal single-particle bases for the light and for the heavy atoms. One can thus use them to enlarge the variational space. In principle, the whole many-body Hilbert space of the $N+1$ problem is spanned by states obtained from $\ket{v}$ by promoting the light atom and/or one or few of the heavy atoms into these excited orbitals. However, one can show that, in the resulting Hamiltonian matrix, the state $\ket{v}$ is directly connected by nonzero matrix elements only with states where the light atom is excited to $\phi_{i>1}$, and a single heavy atom is promoted from $\Psi_{\nu\leq N}$ to $\Psi_{\eta>N}$. Treating these matrix elements as perturbations on top of the diagonal ones, we can calculate the second-order correction to Eq.~(\ref{VarEn}). The result is shown in the left panels of Fig.~\ref{fig2} as dotted curves. We see that, by accounting for pair excitations in this manner, the agreement between the exact and the mean-field approaches improves, also realizing a rather satisfactory cross-check between these theories~\cite{RemThirdOrder}.

In conclusion, we solved the $N+1$ cluster problem in free space, exactly for $N\leq 5$, and by using two mean-field approaches valid asymptotically for $N\gg1$. The Thomas-Fermi density functional approach provides an analytic description of the problem, predicting, in particular, scaling laws for the thresholds and shapes of the clusters. The Hartree-Fock variational method  provides additional information on the cluster structure (parity, for instance) and can be systematically improved. Our findings have implications for theoretical studies of mass and population imbalanced Fermi-Fermi mixtures and for experiments on the ${}^{173}\textrm{Yb}$--${}^{6}\textrm{Li}$ \cite{hara2011,green2020}, ${}^{53}\textrm{Cr}$--${}^{6}\textrm{Li}$~\cite{Neri2020},  ${}^{40}\textrm{K}$--${}^{6}\textrm{Li}$~\cite{Taglieber2008,Wille2008,Voigt2009}, and  ${}^{161}\textrm{Dy}$--${}^{40}\textrm{K}$~\cite{Ravensbergen2018,Ravensbergen2020} mixtures. We also note that our free-space model corresponds to the low-occupation and low-interaction limit of the Fermi-Hubbard model with hopping asymmetry, for which various many-body phases, such as the liquid of $2+1$ trimers, were discussed and studied by using the Monte Carlo, density matrix renormalization group, and bosonization techniques~\cite{Burovski2009,Orso2010,Roux2011,Dalmonte2012}.

We thank M. Zaccanti for fruitful discussions, and acknowledge the support of the ANR grant Droplets (19-CE30-0003-02). DSP is grateful to the Russian Quantum Center for hospitality.


\begin{thebibliography}{50}
\bibitem{remTF} Note that this theory is valid for $N\gg 1$ and does not care if $N$ is an integer or not. Applying it to finite (integer) $N$ leads to artifacts. In particular, clusters with different $N$ always have a finite energy difference even at thresholds.
\bibitem{RemLargeMratio} We have continued the expansion to the next order in $\lambda$ and found that $E_{N+1}- N^2 E_{1+1}\propto (M/m)^{-1/3}$ at large mass ratios (note the slow convergence). Compared with our exact results for $N\leq 5$, we can state that the exponent is captured correctly by this mean-field theory, but the proportionality coefficient is not, although the agreement in determining this coefficient improves with $N$.
\bibitem{RemThirdOrder} We have also calculated the third-order term, which further improves the agreement, but comes with a significantly higher computational cost.
\end{thebibliography}
\end{document}